\documentclass[apjl]{emulateapj}

\shorttitle{Large-scale motions in the Perseus Cluster}
\shortauthors{Simionescu et al.}

\begin{document}

\title{Large-scale motions in the Perseus Galaxy Cluster}

\author{A. Simionescu\altaffilmark{1,2}, N. Werner\altaffilmark{1,2}, O. Urban\altaffilmark{1,2}, and S. W. Allen\altaffilmark{1,2,3}}
\affil{KIPAC, Stanford University, 452 Lomita Mall, Stanford, CA 94305, USA}
\affil{Department of Physics, Stanford University, 382 Via Pueblo Mall, Stanford, CA  94305-4060, USA}
\affil{SLAC National Accelerator Laboratory, 2575 Sand Hill Road, Menlo Park, CA 94025, USA}

\author{A. C. Fabian\altaffilmark{4} and J. S. Sanders\altaffilmark{4}}
\affil{Institute of Astronomy, Madingley Road, Cambridge CB3 0HA, UK}

\author{A. Mantz\altaffilmark{5,6}}
\affil{Kavli Institute for Cosmological Physics, University of Chicago, 5640 South Ellis Avenue, Chicago, IL 60637, USA}
\affil{Department of Astronomy and Astrophysics,University of Chicago, 5640 South Ellis Avenue,Chicago, IL 60637-1433, USA}

\author{P. E. J. Nulsen\altaffilmark{7}}
\affil{Harvard-Smithsonian Center for Astrophysics, 60 Garden St., Cambridge, MA 02138, USA}

\author{Y. Takei\altaffilmark{8}}
\affil{Institute of Space and Astronautical Science (ISAS), JAXA, 3-1-1 Yoshinodai, Chuo-ku, Sagamihara, Kanagawa, 252-5210 Japan}

\begin{abstract}
By combining large-scale mosaics of ROSAT PSPC, XMM-Newton, and Suzaku X-ray observations, we present evidence for large-scale motions in the intracluster medium of the nearby, X-ray bright Perseus Cluster. These motions are suggested by several alternating and interleaved X-ray bright, low-temperature, low-entropy arcs located along the east-west axis, at radii ranging from $\sim10$ kpc to over a Mpc.
Thermodynamic features qualitatively similar to these have previously been observed in the centers of cool core clusters, and were successfully modeled as a consequence of the gas sloshing/swirling motions induced by minor mergers. Our observations indicate that such sloshing/swirling can extend out to larger radii than previously thought, on scales approaching the virial radius.
\end{abstract}


\section{Introduction}

Among the first important {\sl Chandra} results on galaxy clusters was the surprising discovery of cold fronts - remarkably sharp surface brightness edges, with relatively dense, cool gas on the inner (bright) side and low density hot gas on the outer (faint) side \citep{Markevitch2000,Markevitch01}. Cold fronts are seen both in merging systems, where the edge appears at the projected location of the density discontinuity separating the hotter intra-cluster medium (ICM) from the low entropy core of an infalling subcluster, and in the centers of cool-core clusters (at $r=$10--400~kpc) with little or no clear signs of recent mergers. In the latter case, multiple arcs, curved around alternating sides of the central gas density peak along a spiral-like pattern, are often seen \citep[][and references therein]{markevitch2007}. 

Cold fronts in cooling cores are believed to be due to the sloshing (or, more accurately, swirling) of the gas in the gravitational potentials of clusters \citep{Markevitch01}. Numerical simulations show that sloshing can be induced easily by a minor merger, where a subcluster falls in with a nonzero impact parameter \citep{Tittley05, Ascasibar06}. The gas from the subcluster is typically stripped early during the infall, but its dark matter continues to fall in and, due to the gravitational disturbance, the density peak of the main cluster swings on a spiral-like trajectory relative to the center of mass. As the central ICM is displaced, concentric cold fronts are created where cooler and denser parcels of gas from the center come into contact with the hotter outskirts. The cold fronts associated with sloshing/swirling can persist for Gyrs. It has been proposed that the spiral pattern of cold fronts in the innermost regions of cool core clusters signals the presence of large-scale bulk spiral flows \citep{Keshet2011}.

The ICM in the core of the Perseus Cluster of galaxies (A426), which is the brightest, extended extragalactic X-ray source, also shows such a spiral pattern \citep{Fabian00,Churazov00,Churazov03}. A deep 1.4 Ms {\sl Chandra} observation has confirmed the presence of a cold front located about 100~kpc west of the cluster core \citep{Fabian2011}.
Previous Einstein Observatory and {\it ROSAT PSPC} imaging had furthermore suggested the presence of an east-west shift in the centroid of the X-ray isophotes for this system on larger scales of several hundred kpc, which was interpreted as a sign of an ongoing or past merger \citep{BranduardiRaymont81,Schwarz92,Allen92}. In this Letter, we use both archival ROSAT PSPC and XMM-Newton data, as well as a recently obtained Suzaku mosaic of the Perseus Cluster to investigate the origin of these asymmetries in unprecedented detail, from the cluster core out to large radii. We examine possible links between the smaller-scale spiral pattern previously seen in the core and new features reported here on much larger scales, and discuss the implications for the past and current dynamical state of the cluster, and the physical properties of the ICM.

We assume a $\Lambda$CDM cosmology with $\Omega_m$=0.27, $\Omega_\Lambda$=0.73, and H$_0$=70 km/s/Mpc. At the redshift of the cluster, z=0.0179 \citep{Struble99}, one arcminute corresponds to 21.54 kpc.

\section{Observations, data reduction and analysis}

\subsection{ROSAT}

\begin{figure}
\begin{center}
\includegraphics[width=\columnwidth]{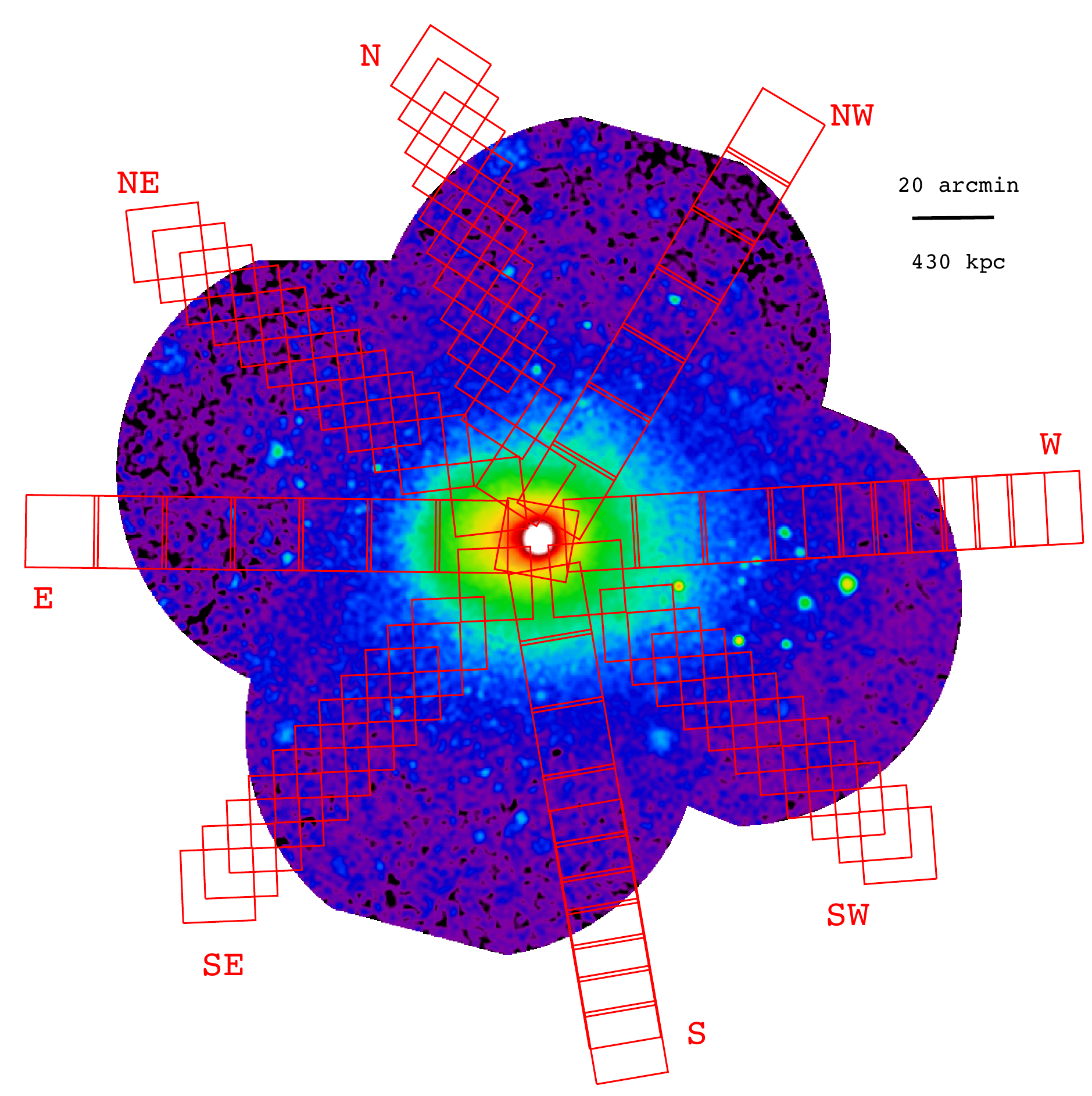}
\end{center}
\caption{X-ray surface brightness map of the Perseus Cluster from the ROSAT PSPC mosaic. Overlaid in red are the locations of the pointings in our Suzaku mosaic. \label{rosat}}
\end{figure}

A mosaic of four off-axis, ROSAT Position Sensitive Proportional Counter (PSPC) observations was performed between February and September 1991, out to radii of $80^\prime$. In addition, we used one PSPC pointing centered on the cluster core and performed in February 1992, which provides better on-axis resolution for the central regions of the cluster. The total exposure time for this ROSAT mosaic is 96~ks.
The data were reduced using the Extended Source Analysis Software (ESAS, \citealt{Snowden1994}). We removed the time intervals of enhanced particle (PB) and scattered solar X-ray (SSX) background. The quiescent PB and SSX components were modeled using the recipes described in \citet{Snowden1994} and subtracted from the data. Background and exposure corrected images in three energy bands (0.7-0.9, 0.9-1.3, 1.3-2.0 keV respectively) were combined, having removed artifacts associated with the detector edges. The result is shown in Fig. \ref{rosat}. The Perseus Cluster appears overall fairly round and relaxed, with a well-formed, bright cool core. 
\vspace{1cm}
\subsection{XMM-Newton}

A mosaic of twelve {\it XMM-Newton} pointings of the Perseus Cluster was taken between 2001--2006, with a total integration time of 374.5~ks. 
Here, we only report results obtained with the EPIC/MOS CCD detectors. 
The filtering and extraction of the MOS data products, the subtraction of their instrumental background, and the exposure correction of the images was performed using the {\it XMM-Newton} Extended Source Analysis Software (XMM-ESAS) and methods described in \citet{kuntz2008} and \citet{snowden2008}. We limit our imaging analysis to the 0.4--7 keV band and only included the central $12^\prime$ of each observation (centered on the aim point) in order to avoid artifacts due to vignetting correction uncertainties. 

\subsection{Suzaku data reduction and analysis}

A large mosaic of observations of the Perseus Cluster was obtained as a Suzaku Key Project during AO 4--6 (2009--2011). This mosaic stretches out to as far as 2 degrees along each of eight arms towards the N, NW, W, SW, S, SE, E, and NE of the cluster center. The location of the pointings is shown in Fig. \ref{rosat}. Here, we focus on the inner $60^\prime$ of this mosaic, where the large-scale features of interest in the surface brightness distribution are most apparent, and where other non-equilibrium effects seen in cluster outskirts are expected to be small.  

The Suzaku data analysis follows that presented in \cite{SimionescuSci,Simionescu_proc}. In addition to the standard screening criteria\footnote{Arida, M., XIS Data Analysis, http://heasarc.gsfc.nasa.gov/docs/suzaku/analysis/abc/node9.html (2010)}, for the XIS1 spectra obtained after the charge injection level increase performed on 2011 June 1, we excluded two columns adjacent on each side to the charge-injected columns (the standard is to exclude one column on either side). This is because the injected charge may leak into these additional columns and cause an increase in the instrumental background.
 
We extracted spectra from annuli centered on the cluster center. The projected and deprojected profiles of thermodynamic properties were obtained with the XSPEC \citep[version 12.5][]{arnaud1996} spectral fitting package, employing the modified C-statistic estimator. We used the {\small projct} model to deproject the data under the assumption of spherical symmetry. Sets of spectra extracted from concentric annuli in the 0.7--7.0 keV band (1.5--7.0 keV band for XIS0) were modeled simultaneously. We modeled each shell as a single-temperature thermal plasma in collisional ionization equilibrium using the {\it apec} code \citep{smith2001}, with the temperature, metallicity, and spectrum normalization as free parameters. The X-ray background was modeled in regions free of cluster emission at radii $r>85^\prime$ with a power-law emission component that accounts for the unresolved population of point sources, two thermal plasma models accounting for the Galactic foreground emission, and one thermal plasma model for the local hot bubble \citep[for details see][]{Simionescu_proc}. The Galactic absorption column density was fixed to the average value measured at the location of each respective Suzaku field from the Leiden-Argentine-Bonn radio HI survey \citep{kalberla2005}.

\section{Results}

\begin{figure*}
\begin{center}
\includegraphics[width=0.9\columnwidth]{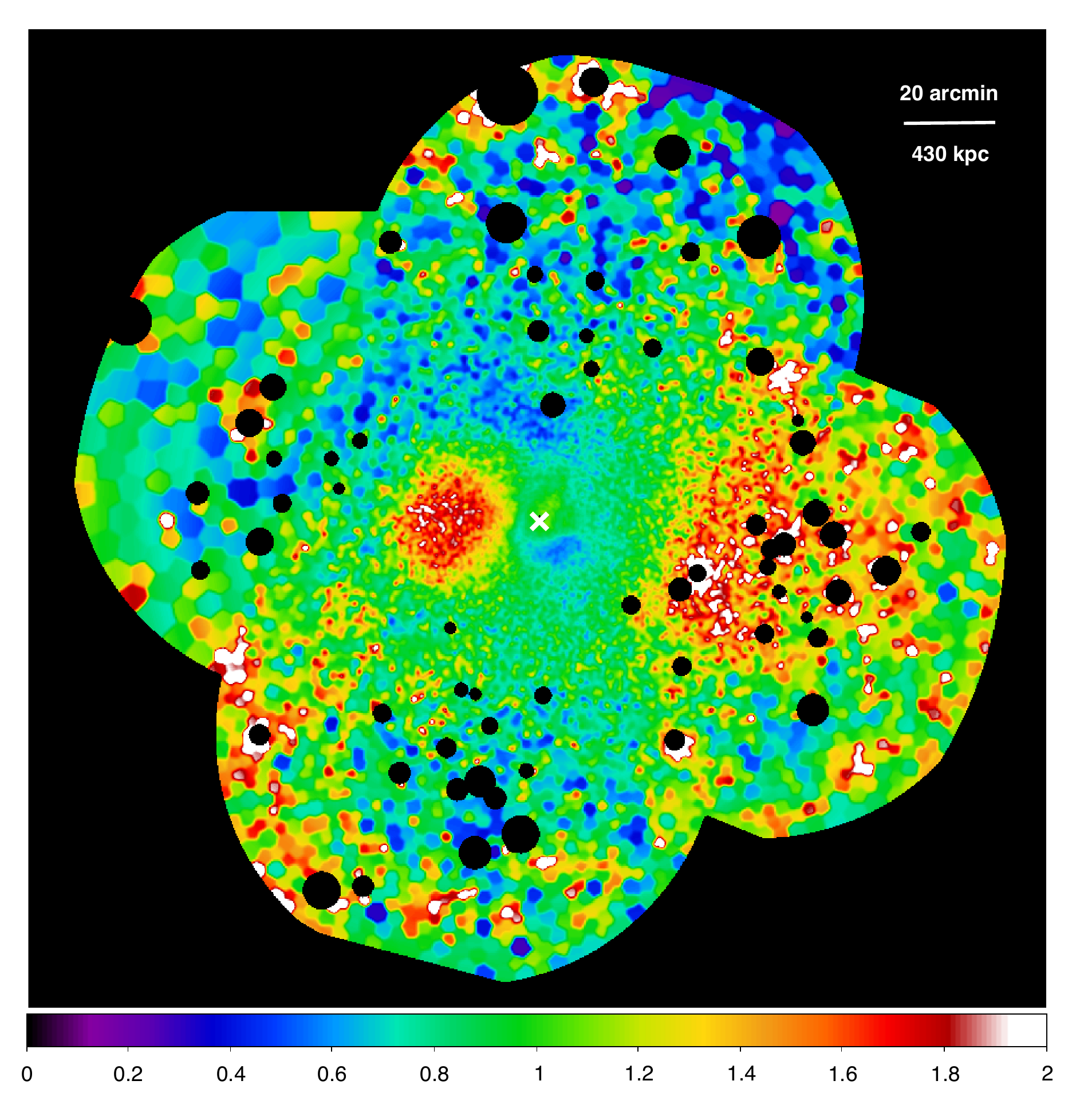}
\includegraphics[width=1.1\columnwidth]{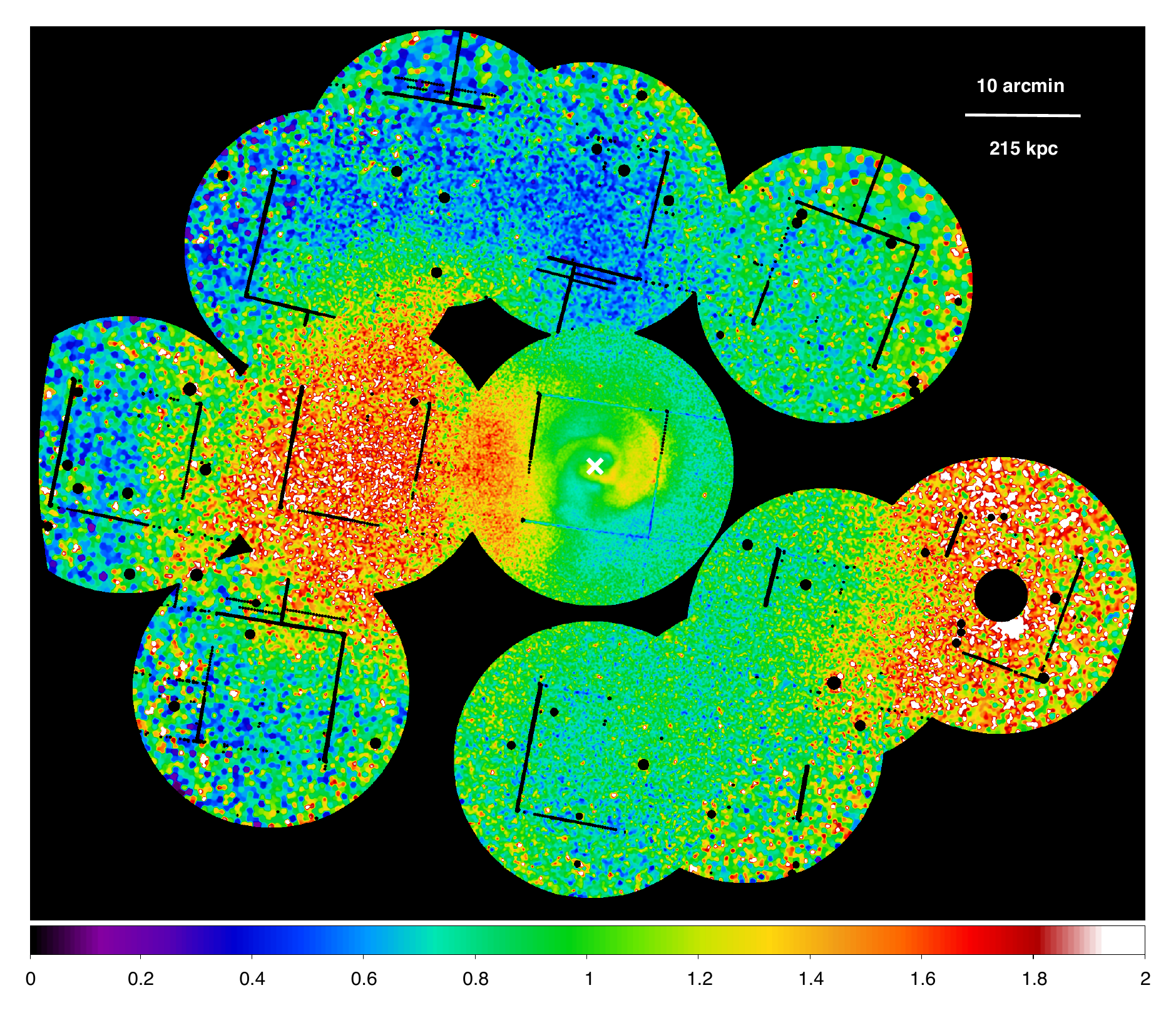}
\end{center}
\caption{Ratio of the X-ray surface brightness distribution with respect to the azimuthal average for the mosaic of ROSAT PSPC observations (left) and XMM-Newton observations (right) of the Perseus Cluster.
Each X-ray mosaic was binned to a minimum of 25 counts per bin using a weighed Voronoi tessellation algorithm before performing the division by the azimuthally averaged radial profile. The black circles represent the point sources that were excluded from the data analysis. The white ``X'' marks the cluster center.
\label{rat}}
\end{figure*}

\begin{figure*}
\begin{center}
\includegraphics[width=\columnwidth]{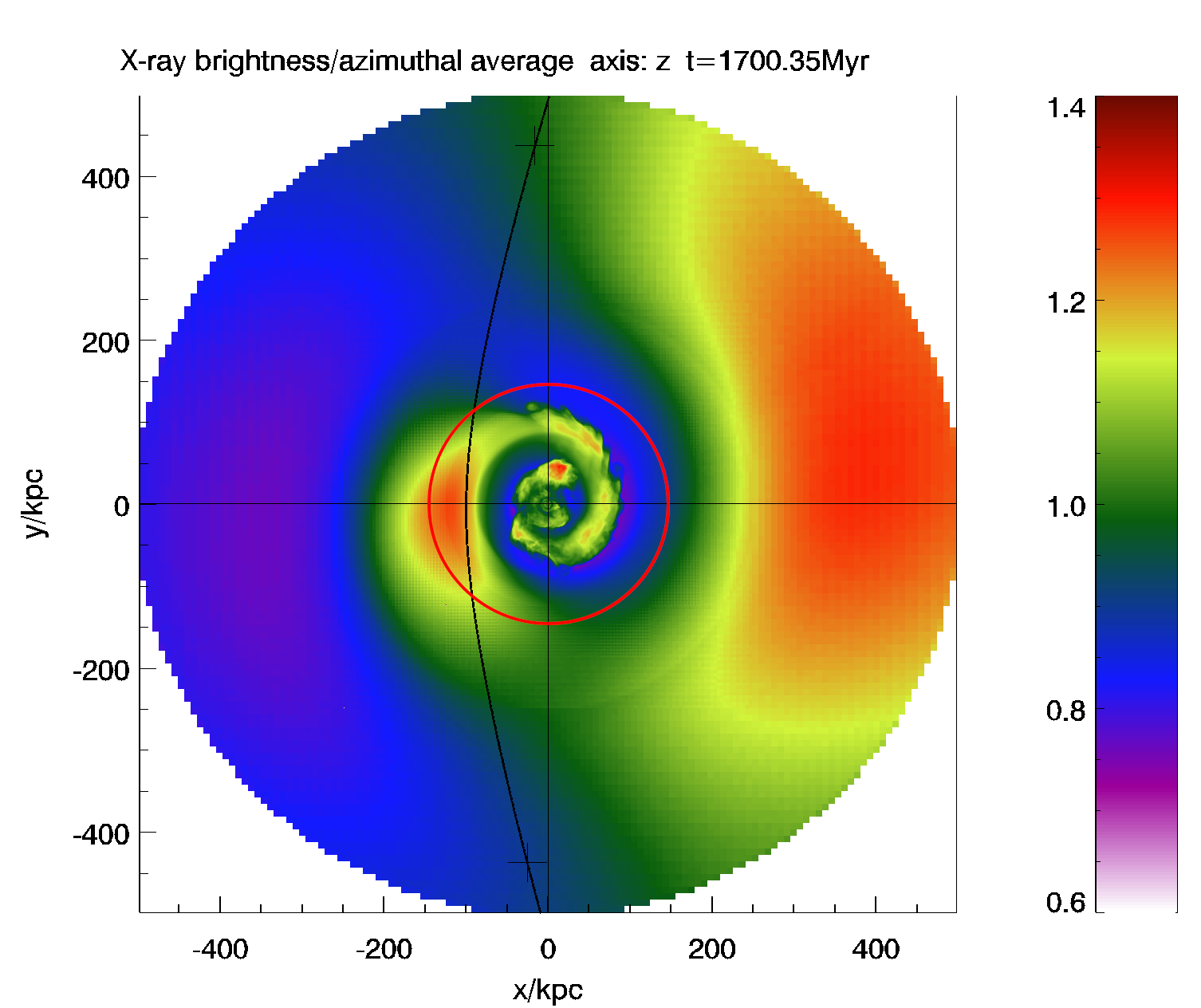}
\includegraphics[width=\columnwidth]{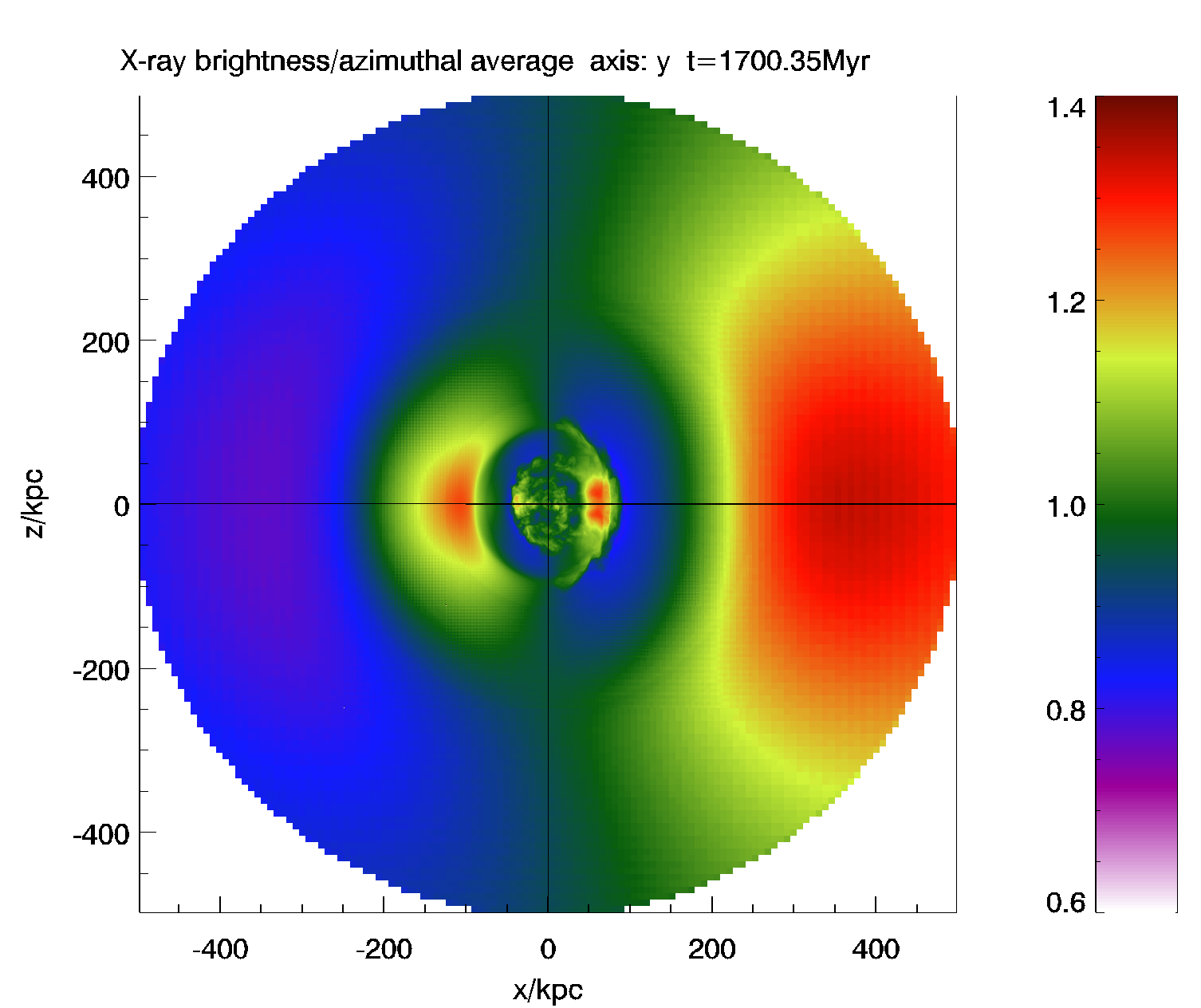}
\end{center}
\caption{Ratio of the simulated X-ray surface brightness distribution with respect to the azimuthal average for numerical simulations of gas sloshing in the Virgo Cluster,  reproduced from \cite{Roediger11}. The left-hand panel shows the plane of the merger face-on; for comparison, we also show the edge-on view on the right-hand side.
The black line in the left panel marks the path of the disturber. Note that the features shown here occur on scales a factor of three smaller than what we observe in the Perseus Cluster. 
\label{sim}}
\end{figure*}

We binned the instrumental-background subtracted, exposure corrected ROSAT and XMM-Newton mosaic images to a minimum of 25 counts per bin using the weighted Voronoi tessellation algorithm presented by \cite{Voronoi06} and divided the resulting maps by the azimuthally averaged radial profile obtained in annuli centered on NGC~1275. For ROSAT, we used 100 annuli with a width of $1^\prime$, while for XMM-Newton we used $5^{\prime\prime}$-wide annuli in the cluster center (inner $500^{\prime\prime}$), where the surface brightness gradient is very steep, and $25^{\prime\prime}$-wide annuli beyond this radius.
Point sources identified with the CIAO task wavdetect, using appropriate sizes for the PSFs of the respective instruments, were excluded from this analysis after having visually rejected spurious detections.
The results are shown in Fig. \ref{rat} and indicate that the X-ray emission of the Perseus Cluster is asymmetric both on small and large scales. 

The ROSAT observations show a clear surface brightness enhancement that extends out to $r\sim700$~kpc from the cluster center towards the east; beyond this radius, we see a clear excess on the opposite side of the cluster towards the west, which reaches all the way to the outer boundary of the ROSAT mosaic. The inner edge of this western outer bright region has a concave shape with respect to the cluster center. We note that dividing by an elliptical double beta model, rather than using the average at each given radius, also reveals the presence of these features. 

Qualitatively, this alternating and interleaved placement of the surface brightness enhancements is remarkably similar to features seen in numerical simulations of gas sloshing induced by minor mergers in cool core clusters. To exemplify this similarity, we show in Fig. \ref{sim} the ratio of X-ray surface brightness with respect to the azimuthal average predicted by the numerical simulations presented in \cite{Roediger11}. Note, however, that, despite the striking qualitative agreement, the features shown in Fig. \ref{sim} occur on scales a factor of three smaller than what we observe in the Perseus Cluster. 

The two panels in Fig. \ref{sim} show two different projections through the simulation volume. In the projection shown on the left, the minor merger that triggered the sloshing motion happened in the plane of the sky; on the right, the merger occurred along the line of sight. The main difference between these two scenarios consists in the presence or absence of a north-south asymmetry, that appears in addition to the main east-west features. In case the merger is seen in the plane of the sky, a bright spiral of almost constant brightness as a function of azimuth is seen in the very center, while on larger scales there is a small surface brightness excess towards the southern side compared to the north. This larger scale excess is, however, on much fainter levels than the contrast of the east-west features. The ROSAT, Suzaku, and XMM-Newton observations indeed all consistently show the presence of such a north-south asymmetry. 

The innermost pointing of the XMM-Newton mosaic (the central $r\sim15^\prime$) shows an over-dense counterclockwise spiral starting at a radius of only 10~kpc, which was discussed in detail by \cite{Churazov03}. This inner spiral is delimited towards the west by a clear cold front located at a radius of about 100~kpc; towards the north of this inner western front, the surface brightness is higher compared to similar radii on the southern side, with the exception of regions associated with cavities inflated by the AGN, where the gas has been pushed out \citep[for more discussion on these cavities see][]{Fabian2011}. This small-level surface brightness enhancement towards the north appears to connect the inner western front to the larger scale eastern excess, which begins just beyond 100~kpc and extends to approximately 700~kpc. 
At radii similar to the outer edge of the eastern excess, the surface brightness to the south of the cluster core is in turn higher than that towards the north; through this enhancement, the eastern brightness excess seems to connect further along a counterclockwise spiral to the outermost western over-dense region that reaches out to radii of well over a megaparsec, approaching the virial radius. Note also that the radius of the concave inner edge of the large-scale western enhancement increases counterclockwise (it is smaller on the southwestern side than on the western side); this is most evident in the ROSAT data, which sample the entire western part of the cluster without gaps in coverage. These features indicate that we are not looking at the sloshing motion exactly edge-on, in which case we would not have expected any north-south asymmetry, as shown in the right panel of Fig. \ref{sim}. Exactly how close the plane of the oscillation is to the plane of the sky cannot be determined without further detailed modeling.

\begin{figure}
\includegraphics[width=1.05\columnwidth]{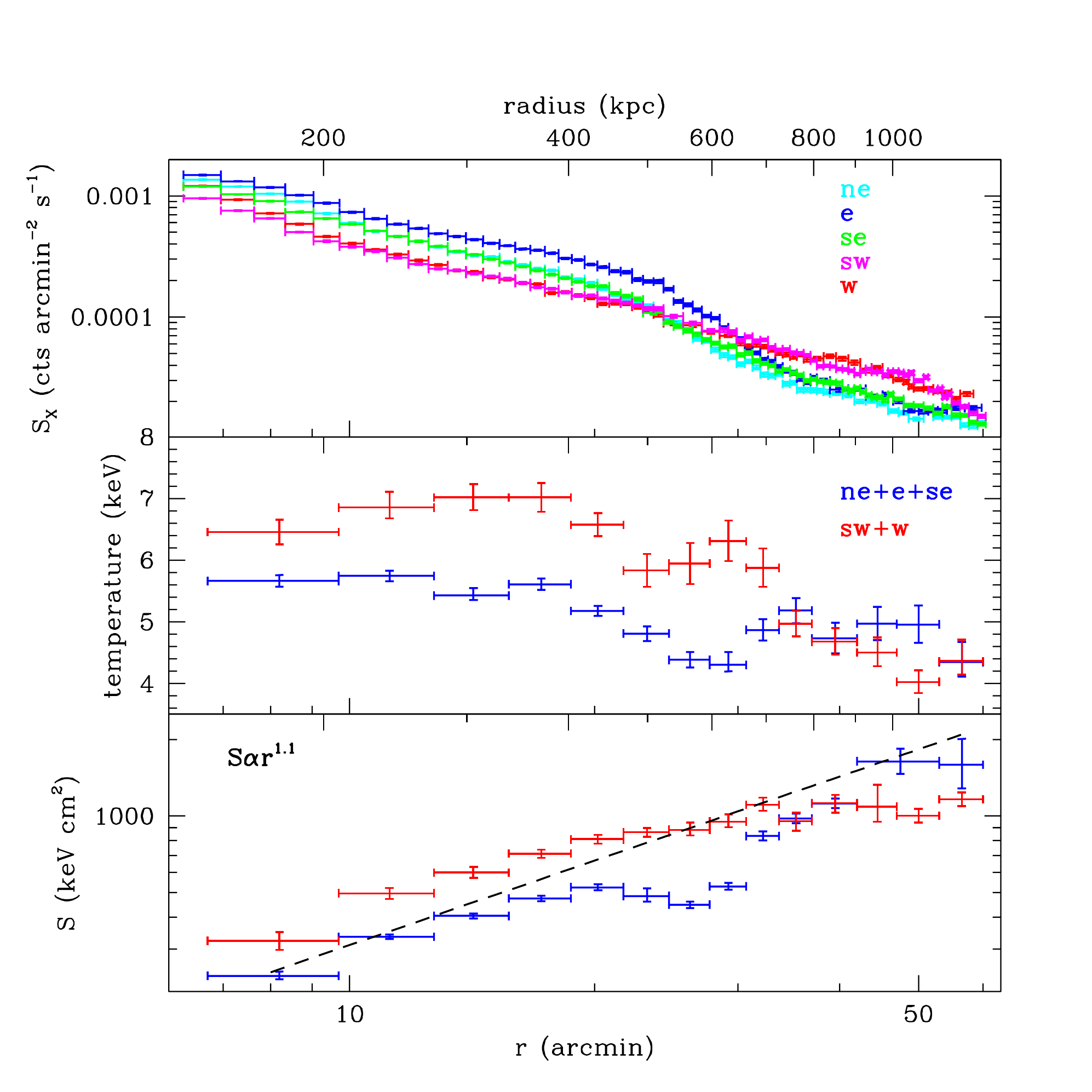}
\caption{Surface brightness, projected temperature, and deprojected entropy profiles obtained from our Suzaku data. The surface brightness profiles are shown individually for the E, SE, and NE directions all sampling the E edge at $\sim700$~kpc, as well as for the W and SW arms showing the large-scale excess beyond this radius. The temperature and entropy profiles were obtained by averaging the three arms towards the E and two towards the W. For reference, in the entropy panel we overplot the best-fit power-law with an index 1.1, fitted to the data from the remaining three relatively more relaxed arms (S,N,NW).
\label{suzaku}}
\end{figure}

The surface brightness profiles obtained from the Suzaku data along several directions that sample either the eastern or the western enhancements are shown in the top panel of Fig.~\ref{suzaku}. 
The surface brightness profiles towards the SE, E, and NE display a clear excess at radii up to $r\sim30^\prime$, and also show quantitatively the variation of this excess as a function of azimuth. Clearly, the highest excess is found towards the east, and a smaller but still highly significant enhancement is seen towards both the SE and NE. We note that, around 30--40$^\prime$, the SE brightness is higher than towards the NE, consistent with the spiral continuing along the southern part of the cluster and joining the excess at larger radii towards the west, that is clearly revealed at radii beyond $r\sim30^\prime$ in the profiles along the SW and W directions. 

Projected temperature profiles averaged along the northeastern, eastern, and southeastern arms of the Suzaku mosaic are presented in the middle panel of Fig.~\ref{suzaku}. These show a lower temperature on the inner, brighter side of the eastern surface brightness edge, consistent with this feature being a cold front as was argued based on the eastern arm data alone in \citet{SimionescuSci}. Inside this discontinuity, the temperature profile towards the east is consistently lower than the temperature profile towards the west. Beyond the discontinuity, the temperature profiles cross, and at large radii the temperature to the west becomes lower than the temperature to the east. 

Deprojected entropy profiles are shown in the lower panel of Fig.~\ref{suzaku}. Including only the effects of gravitational heating as the gas falls into the cluster potential, simple theoretical models and numerical simulations of large scale structure formation predict that the entropy $K$ should follow a power-law form, $K\propto r^{\beta}$, with $\beta\sim1.1 - 1.2$ \citep{tozzi2001,voit2005mnras}. The dashed line shows the best fit power-law with index $\beta=1.1$ to the Suzaku data along the least disturbed  N, NW, S directions at radii $r<50^\prime$. Beyond this radius, gas clumping effects can become significant and affect the measured profiles \citep{SimionescuSci}. The eastern entropy profile follows a power-law like distribution out to a radius of 450~kpc (20$^\prime$), then it flattens, and beyond the surface brightness discontinuity at 700~kpc it recovers the power-law shape. To the west, the entropy profile is higher than to the east and follows a power-law shape out to $r\sim700$~kpc. At larger radii, where the western outer surface brightness enhancement is seen, the entropy flattens and becomes lower than the values measured in the east. The flattening of the entropy profiles indicates that the sloshing/swirling motion has uplifted lower-entropy gas from smaller radii, bringing it to the current location of the observed surface brightness enhancements / cold fronts. The higher entropy observed in the western direction out to $r\sim700$~kpc is driven by the relatively lower gas density at these azimuths, and is most likely the consequence of the ``missing'' gas that has been moved from this sector to different azimuths and larger radii. 
\vspace{0.5cm}
\section{Discussion}

As mentioned above, thermodynamic features qualitatively similar to those presented in this work have been modeled successfully in the centers of cool core clusters. Numerical simulations by \cite{Tittley05} and \cite{Ascasibar06} predict that gas sloshing induced by a disturbance of the cluster's gravitational potential produces precisely this type of opposite-and-staggered features, with alternating regions of high X-ray brightness and low temperature on either side of the main cluster's X-ray peak. 

We note that, based on the presence of the eastern surface brightness edge alone, we would not be able to distinguish whether the eastern excess is caused by a large-scale displacement of the main cluster gas, or whether it originated from the stripped ICM of an infalling subcluster (this was the hypothesis put forth by \citealt{Schwarz92} when first describing this feature). However, given the connection of this feature to both smaller and larger scales as described above, we argue that the large-scale sloshing/swirling interpretation is more likely. The eastern excess both begins - in radius - just beyond the edge of the inner western cold front and ends within the concave inner edge of the outer western surface brightness enhancement. The temperature profiles anticorrelate with the X-ray brightness and gas density, as predicted by numerical models of gas sloshing in cluster cool cores. 

Gas motions in the core of the Perseus Cluster (inner 100~kpc) have already been indicated by the lack of resonant scattering in the 6.7~keV He-like iron line \citep{churazov2004}, which implies a range in velocities of at least half of the sound speed. Significant swirling motion in the ICM may also affect the azimuthal distribution of the different generations of bubbles of relativistic plasma inflated by the central AGN. Future observations with the high resolution X-ray spectrometers on the upcoming Japanese-American Astro-H satellite will allow us to accurately map the line-of-sight component of these gas motions in the cluster core. 

The challenge of the results presented here is to understand how gas sloshing/swirling can reach well beyond the cluster's cool core. Already the cold front at $700$~kpc is well outside the cooling radius (approximately 200~kpc), and thus outside the radial range where the temperature dips in the cluster center. The western surface brightness excess reaches even further in radius, out to well beyond 1~Mpc. 
Numerical simulations have so far only attempted to reproduce sloshing cold fronts within the cool-core region characterized by a steep density gradient and a decrease of the temperature at small radii. Based on these models, \citet{Ascasibar06} conclude that ``with the possible exception of short moments of the sub-cluster flyby generating a conical wake, the cluster stays very symmetric on large scales; the only structure is edges in the center.'' 

\citet{Roediger11} performed a quantitative modeling of gas sloshing in the core of the Virgo Cluster, which exhibits spiral-like features in the surface brightness and gas temperature distributions which are qualitatively similar to those presented here, although on much smaller spatial scales. They found that typical signatures of the oscillation of the main cluster gas do include a large-scale brightness asymmetry. It is interesting to note that this brightness excess, which was also associated with a lower average gas temperature compared to other azimuths, did reach beyond the cooling radius of the Virgo Cluster according to the numerical simulations, out to as far as 400~kpc (see Fig. \ref{sim}).
However, the outermost surface brightness edge that is observed in the Virgo Cluster is only 90~kpc away from its center \citep{Simionescu10}, compared to the location of our eastern edge at $700$~kpc. Moreover, based on Fig. \ref{sim}, it would appear that in order to reproduce the large-scale east-west asymmetries, the preferred trajectory of the infalling subcluster would be oriented roughly north-south; while this cannot be ruled out, it seems much more likely in the case of the Perseus Cluster for a perturber to fall in along the east-west direction, which coincides with the major axis of the cluster, as well as with an elongated chain in the galaxy distribution on even larger scales, as part of the Perseus-Pisces supercluster \citep{Haynes86}. 

It will therefore be important to perform numerical simulations targeted at reproducing the features associated with the gas sloshing on very large scales seen in the Perseus Cluster, in order to understand whether this can be explained by an area of the merger parameter space that has not been modeled before, or whether more complex microphysics such as an enhanced gas viscosity is required for these features to be replicated. Such numerical simulations will also allow us to determine whether the asymmetries seen in Perseus could have been caused by a recent merger, or whether they are due to longer-lived modes of oscillation trapped in the gravitational potential of the cluster. Based on cosmological simulations, the presence of significant turbulence and large-scale bulk motions in the ICM, especially at large radii, is not surprising \citep{Burns10,Nagai07,Rasia06}, but the apparent {\sl coherence} of the large-scale spiral-like structure ranging over at least two decades in radius is unexpected.

The Perseus Cluster has a pronounced cool core, with a sharp peak in X-ray surface brightness, a decreasing temperature towards the center, and obvious signs of ongoing AGN feedback processes \citep[e.g.][]{Boehringer93,Fabian00}. The spiral-shaped brightness enhancement seen in the very center on scales of 10--100~kpc seems to be connected to the larger-scale sloshing, which shows that the motions induced by the merger penetrate the cool core. Therefore, if a relatively powerful merger is required to have induced the large-scale sloshing discussed here, it will be important to understand the circumstances under which this merger does not destroy the cluster's cool core. This analysis may provide more general clues towards the survivability of cluster cool cores during mergers \citep[for a discussion, see][]{Burns08,Million10}. 

\acknowledgments

The authors would like to thank M. Br\"uggen and S. Heinz for helpful discussion. 
Support for this work was provided by NASA through Einstein Postdoctoral Fellowship grant number PF9-00070 awarded by the Chandra X-ray Center, which is operated by the Smithsonian Astrophysical Observatory for NASA under contract NAS8-03060. 
We further acknowledge support from awards NNX12AB64G, NNX10AR48G, NNX09AV64G, and NSF AST-0838187.
The work was supported in part by the U.S. Department of Energy under contract number DE-AC02-76SF00515.
This research is based on observations obtained with XMM-Newton, an ESA science mission with instruments and contributions directly funded by ESA member states and the USA (NASA). The authors thank the Suzaku operation team and Guest Observer Facility, supported by JAXA and NASA. 

{\it Facilities:} \facility{Suzaku}, \facility{ROSAT}.

\bibliography{bibliography,clustersnewest,clustersvirial}

\bibliographystyle{apj}

\end{document}